\documentclass[12pt]{article}
 \usepackage[cp850]{inputenc}
 \usepackage{amssymb,amsmath}
 \usepackage{english}
 \usepackage{parskip}
\newcommand{\be}{\begin{equation}}
\newcommand{\ee}{\end{equation}}
\newcommand{\bea}{\begin{eqnarray}}
\newcommand{\eea}{\end{eqnarray}}
 
 \newcommand{\btau}{\mbox{\boldmath$\tau$}}
 \newcommand{\bpi}{\mbox{\boldmath$\pi$}}
 \newcommand{\brho}{\mbox{\boldmath$\rho$}}
 \newcommand{\bF}{\mbox{\boldmath$F$}}
 
 \title{Compatibility of neutron star masses and hyperon coupling
constants\footnote{Dedicated to Prof. Georg S\"u\ss mann on the occasion of
his 70$^{th}$ birthday}
}
 \author{H.~Huber, M.~K.~Weigel, and F.~Weber
    \\Sektion Physik, Universit\"at
    M\"unchen\\
    Am Coulombwall 1, D-85748 Garching, Germany}
 
\begin{document}
 \baselineskip15pt
 \maketitle

\medskip
PACSnumbers: 97.60.7d, 26.60.+c, 21.80.+a, 21.65.+f
\begin{abstract}
It is shown that the modern equations of state for neutron star matter based
on microscopic calculations of symmetric and asymmetric nuclear matter are
compatible with the lower bound on the maximum neutron-star mass for a
certain range of hyperon coupling constants, which are constrained by the
binding energies of hyperons in symmetric nuclear matter. The hyperons are
included by means of the relativistic Hartree-- or Hartree--Fock
approximation. The obtained couplings are also in satisfactory agreement with
hypernuclei data in the relativistic Hartree scheme. Within the relativistic
Hartree--Fock approximation hypernuclei have not been investigated so far.
\end{abstract}

\clearpage\newpage
Neutron star matter (NSM), bound by gravity, differs from high density
nuclear matter produced in heavy ion reactions in several respects: 1) Since
the repulsive Coulomb force is much stronger than the gravitational
attraction, NSM is much more asymmetric than terrestrial matter in heavy ion
collisions. 2) The weak interaction time scale is small in comparison with
the lifetime of the neutron star (NS), but large in comparison with the
lifetime of high-density matter in heavy ion reactions. For that reasons
dense
matter in high-energy reactions has to obey the constraints of isospin
symmetry and strangeness conservation whereas NSM is a charge neutral system
with no strangeness conservation (see, for instance, Ref.\,\cite{1} with
further references). Extraterrestrial NSM is for that reasons a rather
theoretical object with a very complex structure. Since its density stretches
over an enormous range, reaching from crystalline iron on the surface to
several times of nuclear matter saturation density in the core, one
encounters in the theoretical descriptions many obstacles, for which one has
to introduce theoretical assumptions and extrapolations \cite{1,2}. However
one can try to combine both high density systems in a common theory, which is
possible in modern field theoretical approaches. The first systematic
investigation in this respect was performed by Glendenning \cite{3}, who used
the standard nuclear relativistic Lagrangian extended by inclusion of
hyperons and deltas, i.e.
\begin{equation}
\label{2.1}
\begin{split}
{\cal L}(x) = & \sum_{B=p,n,\Sigma^{\pm^o},\Lambda,\Xi^{o,-,},
  \Delta^{-,o,+,++}}
   {\cal L}^0_B(x) \\
& + \sum_{M=\sigma,\omega,\pi,\rho,\delta}
  \left\{ {\cal L}^0_M(x) + \sum_{B=p,n,\ldots,\Delta} {\cal
L}^{int}_{B,M}(x)
 \right\}
   + \sum_{L=e^-,\mu^-} {\cal L}_L (x) ~.
\end{split}
\end{equation}
where the interaction is mediated by the mesons $\sigma,\omega,\cdots$ and
the leptons are treated as free particles. Furthermore a scalar meson
self-interaction is included. More explicitly the hadron
part is given by:
\begin{equation}
\label{2.7}
\begin{split}
        {\cal L}_H(x) = &  \sum_B \bar\psi_B(x) \left[
        i\gamma^\mu \partial_\mu - m_B + g^B_\sigma \sigma(x) - g^B_\omega
        \gamma^\mu \omega_\mu(x) - f^B_\omega \frac{\sigma^{\mu\nu}}{4m_B} 
        F^\omega_{\mu\nu} \right.
   \\
&\left. -\,\frac{f\pi}{m\pi}\,\gamma^5 \gamma^\mu \btau_B \cdot
\partial_\mu\bpi 
   - \frac{1}{2} g_\rho^B \gamma^\mu \btau \cdot \brho_\mu(x) -
   f_\rho \frac{\sigma^{\mu\nu}}{4m_B} \btau \cdot \bF^\rho_{\mu\nu}\right]
  \psi_B(x)   \\
& +\,\frac{1}{2} \left[\partial_\mu \sigma(x) \partial^\mu \sigma(x) -
  m^2_\sigma \sigma^2(x)\right] +
  \frac{1}{2} \left[ \partial_\mu \bpi(x) \cdot \partial^\mu \bpi(x) -
  m^2_\pi \bpi^2(x)\right] \\
& -\, \frac{1}{4}\bF^\rho_{\mu\nu}(x) \cdot \bF^{\mu\nu,\rho}(x) +
  \frac{1}{2} m^2_\rho \brho^{\mu}(x) \cdot \brho_\mu(x) -
  \frac{1}{4} F^\omega_{\mu\nu}(x) F^{\mu\nu,\omega}(x) +
  \frac{1}{2} m^2_\omega \omega^\mu(x) \omega_\mu(x) \\
& -\,\frac{1}{3} m_N b_N [g_\sigma \sigma(x)]^3 - \frac{1}{4} c_N [g_\sigma
\sigma(x)]^4 ~,
\end{split}
\end{equation}
with
\begin{equation}
\label{2.8}
 F^\omega_{\mu\nu}(x)
   \equiv \partial_\mu \omega_\nu(x) - \partial_\nu \omega_\mu(x) ~, \quad
\bF^\rho_{\mu\nu}(x) \equiv \partial_\mu \brho_\nu(x) - \partial_\nu
\brho_\nu(x)~. 
\end{equation}
Within this scheme the vast majority of investigations was performed using
coupling parameters in the nucleonic sector adjusted to the nuclear matter
parameter ($\rho_o,$ E/A, K, J) and the Dirac mass $\tilde m$ and a variety
of
choices for the hyperon couplings (see, for instance, Refs.\,\cite{1,3,4,5}).
The calculations for such Lagrangians could be performed without numerical
convergence problems and resulted in a high strangeness contents of NSM
\cite{1,3}. However it was pointed out in two more recent investigations,
that due to the selected high Dirac mass ($\simeq~0.79~m_N$) such
parametrizations are neither in accordance with standard phenomenological
parametrizations \cite{6}, as, for instance, the frequently employed
parametrizations NL1 and NL-SH, or with parametrizations adjusted to the
outcome of microscopic relativistic Brueckner--Hartree--Fock calculations
(RBHF), respectively, which give a lower Dirac mass of approximately
$0.6~m_N$ \cite{7,8} (necessary also, for instance, for a correct
reproduction of the spin-orbit-splitting \cite{6}). Attempts to use directly
such parametrizations in NSM--calculations, which one should prefer
theoretically, have not been successful, since they lead to negative nucleon
Dirac masses by inclusion of hyperons \cite{1,7}. This strong decrease of
$\tilde m$, caused by the stronger $\sigma$--field in the case of lower Dirac
masses at saturation in comparison to the high choice of $\tilde m$ by
Glendenning, seems to be also not in accordance with the behaviour of $\tilde
m$ in RBHF--calculations for higher densities in symmetric nuclear matter
\cite{9}. For that reason one has to correct this feature of the RH, which
occurs also in the relativistic Hartree--Fock--approximation, by a slight
change of the Lagrangian for higher densities, which gives a more moderate
decrease in this region similiar to the RBHF--values but does not change the
equation of state (EOS) in the region up to moderate densities above
saturation nuclear density \cite{1}. Within this scheme, described in detail 
in Ref.\,\cite{1}, where the changes in
the masses are moderate and the high--density behaviour is through the
dominance of the $\omega$--repulsion hardly altered,  one obtains now in the
RH--approximation NS--matter
compositions with a high strangeness component which are qualitatively
similar \cite{1} to former investigations, but have now the advantage, that
they are based either on a correct phenomenological description of nuclear
properties and the outcome of microscopic RBHF--calculations, with an
additional assumption, done in a controlled and changeable manner by one
additional parameter. A closer inspection reveals that all the other previous
attempts, where such problems did not arise at a first glance, have made more
or less implicitly, for instance, by selecting a high Diracmass \cite{3,5} or
by introducing a $\omega$ self-interaction \cite{10}, such an additional
assumption, however with a rather specific enforcement of the high density
behaviour on the Dirac mass (for details, see Ref.\,\cite{1}). In the
RH--scheme, $\Delta$'s usually do not occur, since due to the large
$\rho$--coupling, needed to reproduce the correct symmetry coefficients, the
charge-favoured but isospin-unfavoured $\Delta^-$ is suppressed \cite{1,3}.
Since the data from the hypernuclei are not yet conclusive  with respect to
the hyperon couplings (for more details, see Refs.\,\cite{11,12}), it is
common in more modern investigations to use the SU(6) symmetry for the vector
couplings and adjust the $\sigma$--coupling with respect to the lowest
hyperon level in nuclear matter. However the data from hypernuclei still
permit a large bandwidth for the couplings \cite{1,6,11,12,13}. As it was
pointed out by Glendenning and Moszkowski \cite{13} one can also use the
maximum neutron-star mass of at least 1.5 solar mass, the so-called
Oppenheimer--Volkoff limit M$_{OV}$, as an additional constraint besides the
binding energy of the hyperon in matter to fix the permitted pairs of
coupling constants $x_\sigma = g_{\sigma H}/g_{\sigma N} (<1)$ and $x_\omega
= g_{\omega H}/g_{\omega N}(\le 1)$. However one has to keep in mind that
such fixations depend strongly on the chosen nucleonic parametrization in
combination with the selected approximation \cite{1}. It seems therefore
worthwhile to revisit this problem in a short note in view of the discussed
theoretically more satisfying parametrizations, based on the outcome of
RHBH--calculations with the modern Brockmann--Machleidt OBE-potentials
\cite{1,7,14}. The parametrizations are given in Table\,I and Table\,II.

More subtle is the case of the RHF--approximation \cite{4,14}, where the
constraint of the hyperon binding energy in nuclear matter is difficult to
implement \cite{1}. A further new feature is that due to the exchange
contributions the  $g_\rho$--coupling becomes smaller in the direction
of the values of the realistic OBE--potentials with the net effect that now
the $\Delta$'s can play an important role in the composition of NSM, so
decreasing the strangeness contents \cite{1}. The $\Delta$--component is
rather sensitive to the choice of the $\Delta$--coupling, and one obtains
qualitatively rather different compositions, reaching from the standard
picture for lower $\Delta$--couplings, as recommended by Rapp et al.
\cite{15}, to $\Delta$--dominated compositions with much smaller strangeness
obtained for $g_\Delta/g_N = 1$ \cite{1}. Since a final solid statement about
the choice of these couplings cannot be made, one cannot exclude, at present,
the possibility that the composition of NSM might deviate from the standard
picture of a high strangeness component, which may be a special feature of
the RH--approximation with a large $\rho$--coupling (for details, see,
Ref.\,\cite{1}). The chosen parametrizations are given in Table\,III. The
dependence of the relative scalar coupling for the hyperons for a selected
relative vector coupling is exhibited in Fig.\,1. As far as the
RH--approximation is concerned all parametrizations with $x_\omega \geq 0.5$
give maximum masses above $1.6~M_\odot$, the lowest maximum gravitational
masses are obtained for no coupling, i.e. RH6, where one obtains $M_{OV} =
1.17~M_{OV}$. The Oppenheimer--Volkoff masses  for the SU(6)--values are:
$M_{OV} = 1.68~M_\odot$ with a baryonic mass of $M_{OVB} = 1.92~M_\odot$ for
nonrotating NSs and $M_{OV}=2.03~M_\odot (M_{OVB} = 2.32~M_\odot)$ for NSs
rotating with their Kepler frequency $\Omega_K = 7.97\times 10^3 s^{-1}$. The
OV--masses increase with stronger hyperon-coupling to approximately
$2.15~M_\odot (\Omega = 0)$ or $2.45~M_\odot (\Omega = \Omega_K)$,
respectively, for the EOS--RH4. In this case with the strongest hyperon
interaction one obtains the largest hyperon contribution, which is shown in
Fig.\,2.  Since the masses of most of the pulsars are
in the range of $1.5~M_\odot$ (the most accurately measured mass is that of
the PSR 1913+16 with $M/M_\odot = 1.422\pm0.003$ \cite{16}, all pairs of
hyperon couplings seem to be compatible or at least not in contradiction with
the observations. The comparison for the RH is depicted in Fig.\,3. In the
framework of the RHF--approximation \cite{1,4,14} the situation is more
complicated. Naive use of the hyperon couplings from the RH--treatment is not
compatible with the binding energies of hyperons in saturated nuclear matter,
since in this case the exchange contributions contribute strongly different
in the time-like and scalar parts of the self-energies and hence cause an
overestimation of the $\Lambda$--binding energy in matter. In this case the
choice with SU(6) vector couplings is not sufficient to reach a OV--mass of
$1.5~M_\odot\,(M_{OV} = 1.35~M_\odot)$. By correcting this feature one
obtains
lower scalar couplings (see Table\,III) in comparison with the RH. The
OV--masses corresponding to the SU(6) $\Lambda$--coupling, i.e. RHF3, give
now $M_{OV}= 1.59~M_\odot\,(M_{OVB} = 1.86~M_\odot)$ for a static,
nonrotating
spherical NS and $M_{OV} = 1.88\,(M_{OVB} = 2.20~M_\odot)$ for the deformed
NS
rotating at Kepler frequency $\Omega_K = 10.51\times 10^3 s^{-1}$. For this
EOS the $\Delta$s contribute significantly, since the strong coupling of the
$\Delta$s in combination with the weaker $\rho$--coupling is capable to
overcome their iso-spin unfavoured deficit. However decrease of the
$\Delta$--coupling according to Ref.\,\cite{15} is sufficient to push out
again the iso-spin unfavoured $\Delta$s and the hyperons contribute in a
similiar manner as in the RH. The results for the OV--masses (RHF9) are then
in the same range, namely: $M_{OV} = 1.65~M_\odot\,(M_{OVB} = 1.89~M_\odot)$
for the static NS and $M_{OV} = 1.94~M_\odot\,(M_{OVB} = 2.22~M_\odot)$ for
the rotating star with the Kepler frequency $\Omega_K = 8.26\times 10^3
s^{-1}$. The outcome for the RHF is depicted in Fig.\,3, too. From figures 1
and 3 one can infer that the parametrizations with $0.5\leq
x_{\omega\Lambda} \leq 0.8$ for the RH-- and RHF--approximations are with
respect to the hyperon couplings in accordance with the NS--mass data, since
they are capable to sustain a mass above $1.5~M_\odot$, which is in
accordance with the observation of the pulsar PSR 1913-16 with a mass of
$1.44~M_\odot$. In this context one should remark that for an isolated NS the
maximum mass is smaller than the OV--mass emerging from the calculation of a
cold NS. This difference is caused by the fact that NSs at birth are composed
of so-called supernova matter with a high lepton fraction characterized by
adjusted to the outcome of our RBHF--calculations one needs a slightly lower
$x_{\sigma\Lambda} = 0.59$ in order to obtain the correct lowest binding
energy of $\Lambda$ in $^{17}O$, \cite{20}. In view of the included
density-dependence such deviations seem reasonable \cite{20}. For the
RHF--approximation a calculation of $\Lambda$--hypernuclei has not been
performed so far.

In conclusion we have determined a range of possible EOSs for NSM, based on
microscopic EOSs of symmetric and asymmetric nuclear matter, not possessing
the unjustified high Dirac masses of earlier treatments and which are
mutually compatible with neutron-star masses, $\Lambda$--binding in nuclear
matter
and hypernuclear levels (in the RH). The relation between the scalar and
vector coupling was fixed by the binding energy, for instance, of the lowest
$\Lambda$--level in nuclear matter by the Hugenholtz--Van Hove theorem.
So far the comparison has been only tested for the relativistic
Hartree--approximation, since hypernuclei calculations in the
RHF--approximation are not available at present.

{\bf Acknowledgements}
We thank Ch. Schaab for valuable discussions, H. Huber acknowledges support
from a grant of the Bavarian State.

\clearpage\newpage
\begin{table}
 \begin{center}
 TABLE I \\[1cm]     
\begin{tabular}
{|l|r|r|r|r|r|c|}
\hline
 & $g_\sigma$~~ & $g_\omega$~~ & $g_\rho$~~ & $10^3 \times b_N$ & $10^3
\times c_N$ & $f_\rho/g_\rho$ \\
\hline
RHA & 9.58096 & 10.67698 & 3.81003 & 3.333665 & -3.52365 & -- \\
RHF\,A1  & 9.28353 & 8.37378 & 2.10082 & 3.333689 & -2.15239 & -- \\
RHF\,A2 & 9.24268 & 8.25548 & 2.19809 & 2.96514 & -2.68614 & 3.7 \\
RHF\,A3 & 9.16665 & 8.07540 & 2.37987 & 1.95524 & -2.36335 & 6.6 \\
RHB   &  9.59169 & 10.68084 & 3.66541 & 3.62616 & -4.17140 & -- \\
RHF\,B1 & 9.36839 & 8.40466 & 1.77326 & 3.74354 & -3.18456 & -- \\
RHF\,B2 & 9.33266 & 8.32154 & 1.86078 & 3.44306 & -3.46261 & 3.7 \\
RHF\,B3 & 9.26782 & 8.19391 & 2.02216 & 2.67292 & -3.18198 & 6.6 \\
\hline
\end{tabular}
\end{center}
\end{table}

\begin{table}
\begin{center}
TABLE II \\[1cm]    
\begin{tabular}{|l|l|l|l|l|l|l|}
\hline
Approximation & $x_{\sigma\Sigma\Lambda}$ & $x_{\omega\Sigma\Lambda}$ &
      $x_{\sigma\Xi}$ & $x_{\omega\Xi}$ & $x_{\sigma\Delta}$ &
$x_{\omega\Delta}$\\
\hline
\hline
RH1(SU(6)) & 0.611 & 2/3 & 0.346 & 1/3 & 1 & 1 \\
\hline
RH2 & 0.40 &  0.40 & 0.40 & 0.40 & 1 & 1 \\
\hline
RH3 & 0.64 & 0.7 & 0.63 & 0.7 & 1 & 1 \\
\hline
RH4 & 0.79 & 0.9 & 0.79 & 0.9 & 1 & 1 \\
\hline
RH5  & 1 & 1 & 1 & 1 & 1 & 1 \\
\hline
RH6 & 0 & 0 & 0 & 0 & 0 & 0 \\
\hline
\end{tabular}
\end{center}
\end{table}

\begin{table}
\begin{center}
TABLE III \\[1cm]
\begin{tabular}{|c|c|c|c|c|c|c|}
\hline
Approximation & $x_{\sigma\Sigma\Lambda}$ & $x_{\omega\Sigma\Lambda}$ &
      $x_{\sigma\Xi}$ & $x_{\omega\Xi}$ & $x_{\sigma\Delta}$ &
$x_{\omega\Delta}$\\
\hline
\hline
RHF1 & 0.611 & 2/3 & 0.346 & 1/3 & 1 & 1 \\
\hline
RHF2 & 0.79 &  0.9 & 0.79 & 0.9 & 1 & 1 \\
\hline
RHF3 & 0.44 & 2/3 & 0.26 & 1/3 & 1 & 1 \\
\hline
RHF4 & 0.36 & 0.5 & 0.35 & 0.50 & 1 & 1 \\
\hline
RHF5  & 0.46 & 0.7 & 0.45 & 0.7 & 1 & 1 \\
\hline
RHF6 & 0.56 & 0.9 & 0.55 & 0.9 & 1 & 1 \\
\hline
RHF7 & 1 & 1 & 1 & 1 & 1 & 1 \\
\hline
RHF8 & 0.44 & 2/3 & 0.26 & 1/3 & 0.625 & 0.625 \\
\hline
RHF9 & 0.44 & 2/3 & 0.26 & 1/3 & 0.625 & 1 \\
\hline
RHF10  & 0.44 & 2/3 & 0.26 & 1/3 & 0.75 & 1 \\
\hline
RHF11  & 0.44 & 2/3 & 0.26 & 1/3 & 0 & 0 \\
\hline
\end{tabular}
\end{center}
\end{table}

\clearpage\newpage
\begin{center}
{\bf Table captions}
\end{center}
\begin{enumerate}
\item[Table I:]  
Parametrizations of the RH-- and RHF--Lagrangian adjusted
to the RBHF--calculations.  For the masses the following values were selected
(MeV):
$m_N = 939$, $m_\sigma = 550$, $m_\omega = 738$, $m_\pi = 138$, $m_\rho 
= 770\,(g_\pi = 1.00265,~f^2_\pi/4\pi = 0.08$). The parametrizations are 
labelled as follows: RHA
$\hat =$ RBHA etc. for the potential $A$ (effective mass at the Fermi
surface $\tilde m = 617.8$~MeV\,($A$); 621.8~MeV~($B$)).

\item[Table\,II:]  
Relative coupling strengths of the hyperons in the relativistic Hartree
approximation. 
In Ref.\,\cite{1} occured a misprint for $x_{\sigma\Lambda}(=0.676)$ and
$x_{\sigma\Xi} (=0.342)$.
For RH1 -- RH4  $x_{\sigma H}$ was
 adjusted for a selected $x_{\omega H}$ according to the
Van--Hove--Hugenholtz theorem, which gives for the
hyperon potential depth $U(H) = x_{\sigma H} \Sigma_s + x_{\omega H}
\Sigma_o$, where $\Sigma_s$ and $\Sigma_o$ are the scalar and vector part of
the self-energy at saturation. For comparison we included also the universal
coupling (RH5) and the free Fermi gas approximation for the hyperons (RH6).
For the nucleonic sector the parametrization RHB was used.

\item[Table\,III:]
Relative coupling strengths of the hyperons in the relativistic Hartree--Fock
approximation. In RHF1 and RHF2 the coupling strengths of the
RH--approximation were used. For RHF3 -- RHF11 $x_\sigma$ was adjusted for a
given $x_\omega$ according to the binding energy of the hyperon in symmetric
nuclear matter at saturation. For the nucleonic sector the parametrization
RHF\,B1 was used. RHF10 is characterized by fixing the potential depth
of $\Delta^-$ to the same value as in RH1 (for the composition, see,
Ref.\,\cite{1}).
\end{enumerate}

\begin{center}
{\bf Figure captions}

\begin{enumerate}
\item[Fig.1.]
Constraint of the relative $\sigma$--coupling constant for a given
$\omega$--coupling due to the lowest hyperon binding energy in saturated
nuclear matter for the RH-- and RHF--approximation, respectively.
\item[Fig.2.]
Baryon-lepton composition as function of the density of a NS in the
RH--approximation with the strongest hyperon--coupling (RH4).
\item[Fig.3.]
Oppenheimer--Volkoff masses for static neutron stars for the different
approximations as function of the relative $x_{\omega H}$--coupling.
$x_{\sigma H}$ is adjusted for these parametrization according to the lowest
hyperon binding energy in matter. The horizontal line gives the mass of the
pulsar PSR 1913+16. $x_{\omega H} = 0.8\,(x_{\sigma H} < 0.72)$ gives the
highest value for the coupling constant compatible with properties of
hypernuclei \cite{12,20}.
\end{enumerate}

\end{center}

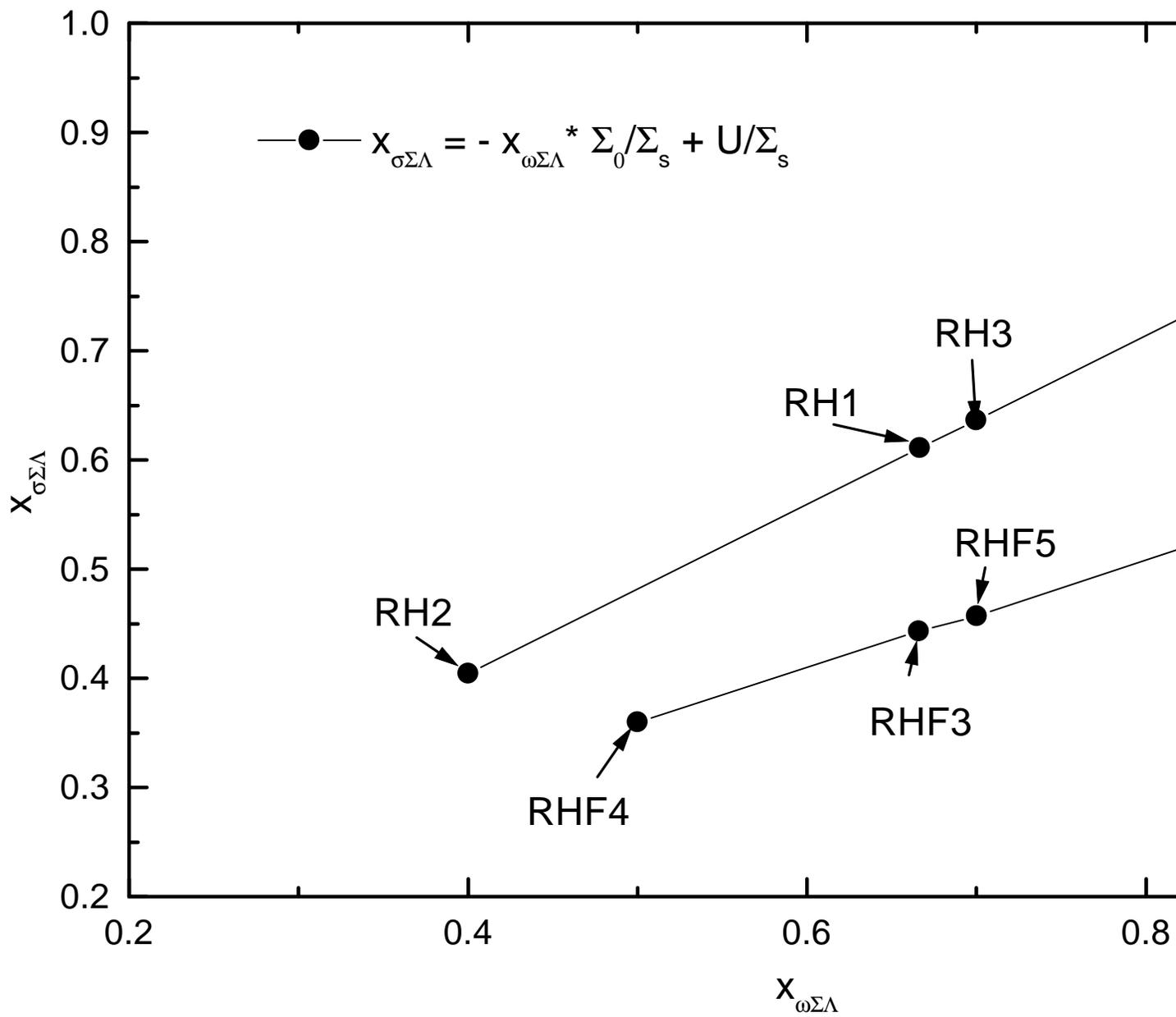
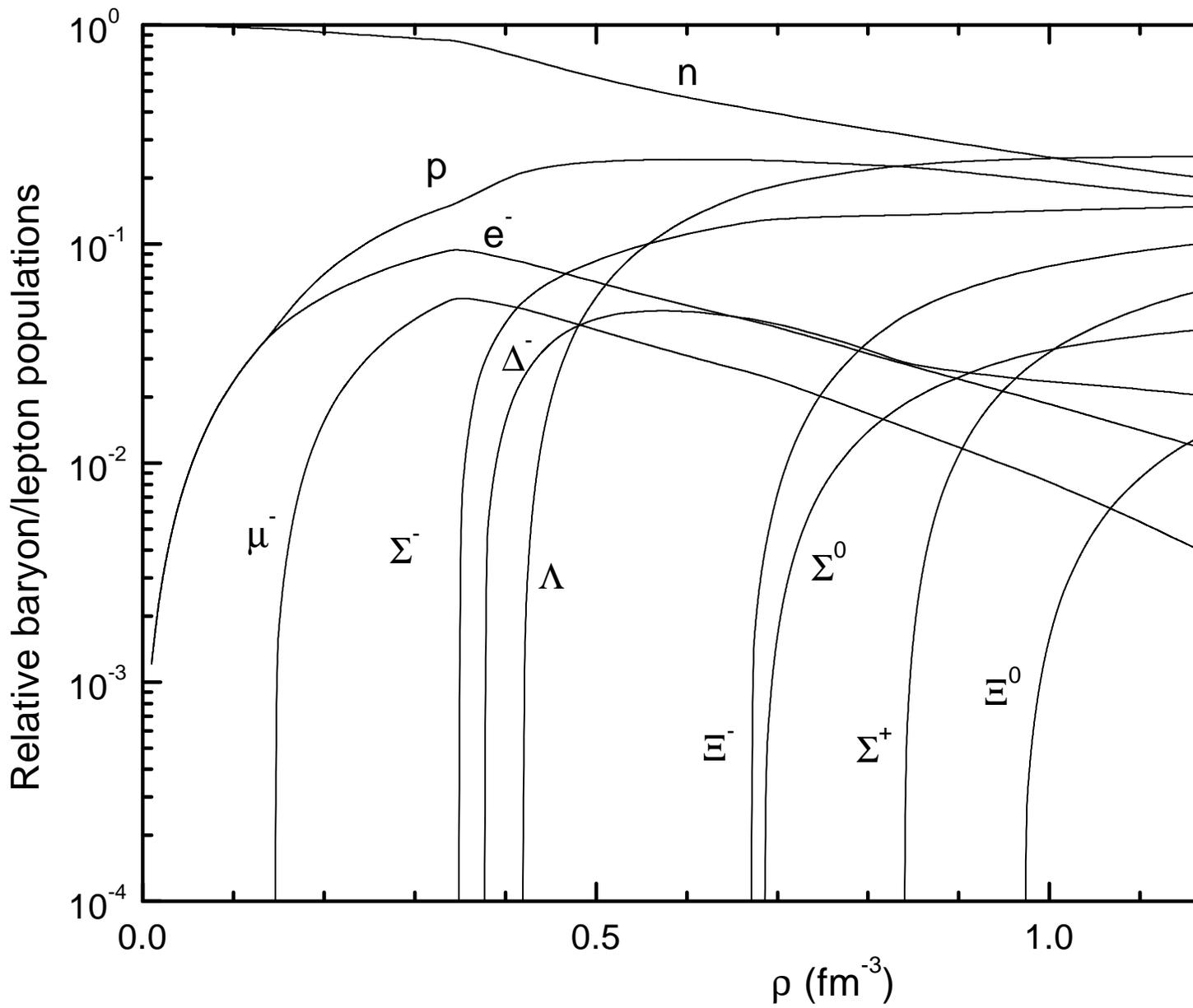
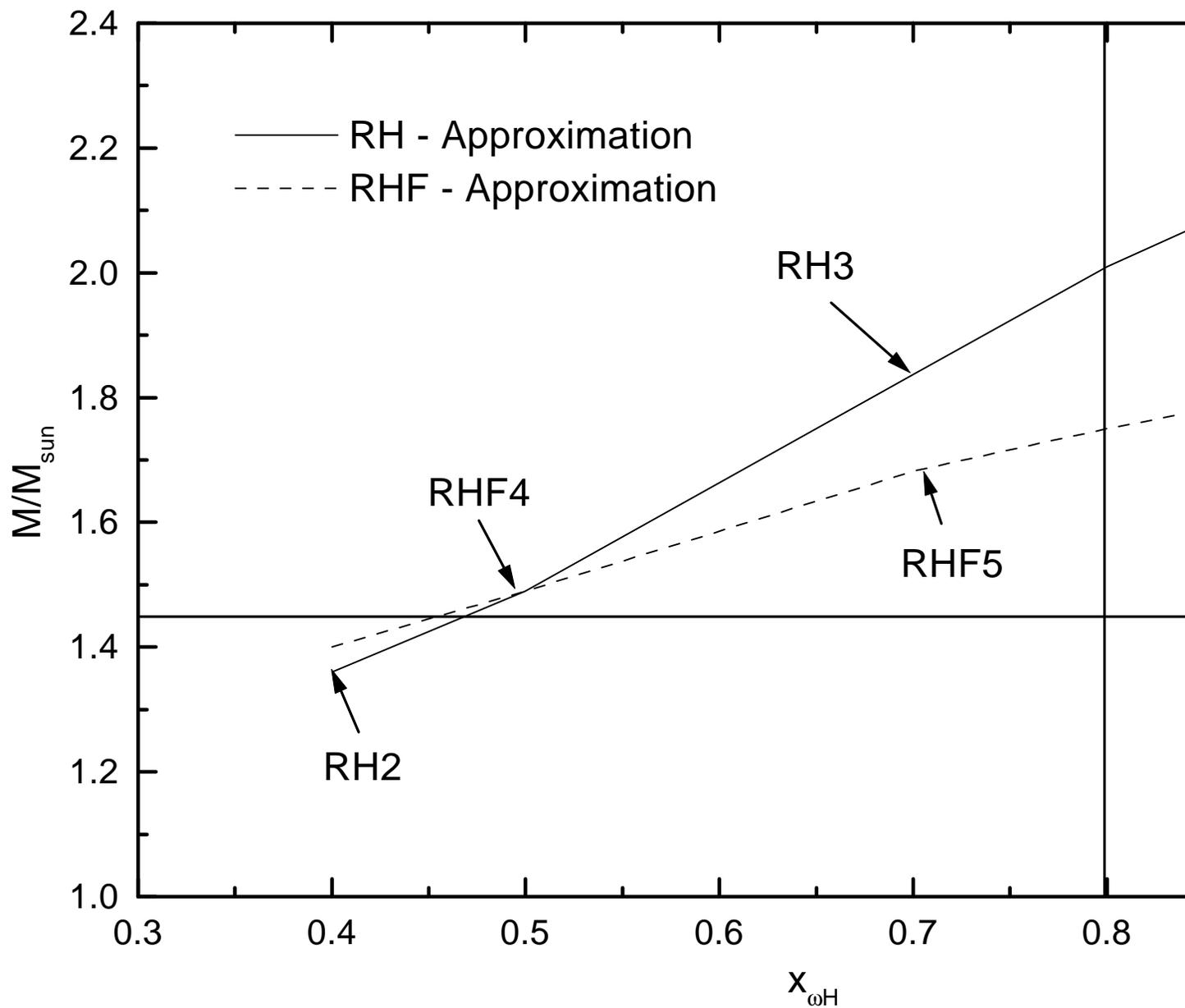
\end{document}